\documentclass[]{bytedance_seed}

\usepackage[toc,page,header]{appendix}

\usepackage{minitoc}
\usepackage{multirow}
\usepackage{array}
\usepackage{threeparttable}
\usepackage{subcaption}
\usepackage{float}
\usepackage{tabularx}
\usepackage{amssymb}
\usepackage{amsmath}

\title{ AsyncHZP: Hierarchical ZeRO Parallelism with Asynchronous Scheduling for Scalable LLM Training  }

\author[1,*]{Huawei Bai}
\author[1,*]{Yifan Huang}
\author[1,*]{Wenqi Shi}
\author[1,*,\dagger]{Ansheng You}
\author[]{Feifan Shao}
\author[1]{Tengfei Han}
\author[1,\dagger]{Minghui Yu}

\affiliation[1]{ByteDance Seed}

\contribution[*]{Equal Contribution}
\contribution[\dagger]{Corresponding authors}

\abstract{

The training efficiency and scalability of language models on massive clusters currently remain a critical bottleneck. Mainstream approaches like ND parallelism are often cumbersome and complex, while flexible alternatives such as the Zero Redundancy Optimizer (ZeRO) are frequently hampered by communication overhead. In this paper, we propose Asynchronous Hierarchical Zero Parallelism (AsyncHZP), a novel asynchronous variant of ZeRO designed to achieve superior performance while maintaining simplicity and memory efficiency. Unlike traditional ZeRO, which employs over-fine-grained sharding that can lead to inefficient communication, AsyncHZP adaptively reshards parameters, gradients, and optimizer states across different replica groups. This strategy optimizes device memory utilization and significantly reduces communication overhead. In addition, we also design a multi-stream asynchronous scheduling method that executes parameter all-gather and gradient reduce-scatter operations in dedicated background threads, effectively overlapping communication with computation while incurring negligible memory fragmentation. Empirical evaluations on both Dense and Mixture-of-Experts (MoE) models confirm that AsyncHZP maintains robust stability at scale. It consistently outperforms classic ND parallelism, achieving state-of-the-art performance without complex strategic tuning, thereby simplifying the path to efficient large-scale training.
}

\date{October 20, 2025}
\correspondence{\email{youansheng@bytedance.com},  \email{yuminghui.exp@bytedance.com}}

\begin{document}
\maketitle

\section{Introduction}

\label{sec:intro}

Recently, deep learning models have demonstrated exceptional efficacy across diverse domains such as natural language processing, image/video analysis, speech recognition, etc.
Based on the Transformer\cite{DBLP:journals/corr/attention} architecture, larger models typically yield better training performance, driving a steady growth in model size from billions to hundreds of billions of parameters and beyond.
However, this rapid growth has amplified training challenges, including increasing memory usage, higher computational burdens, and greater complexity in parallelization strategies. To address these challenges, two predominant approaches have emerged for facilitating training across large-scale distributed systems: ND parallelism \cite{megatron-2021,deepspeed3dblog} and the Zero Redundancy Optimizer (ZeRO) \cite{rajbhandari2020zero, rajbhandari2021zero, zeroplusplus}.

\begin{figure}
    \centering
    \includegraphics[width=0.75\columnwidth]{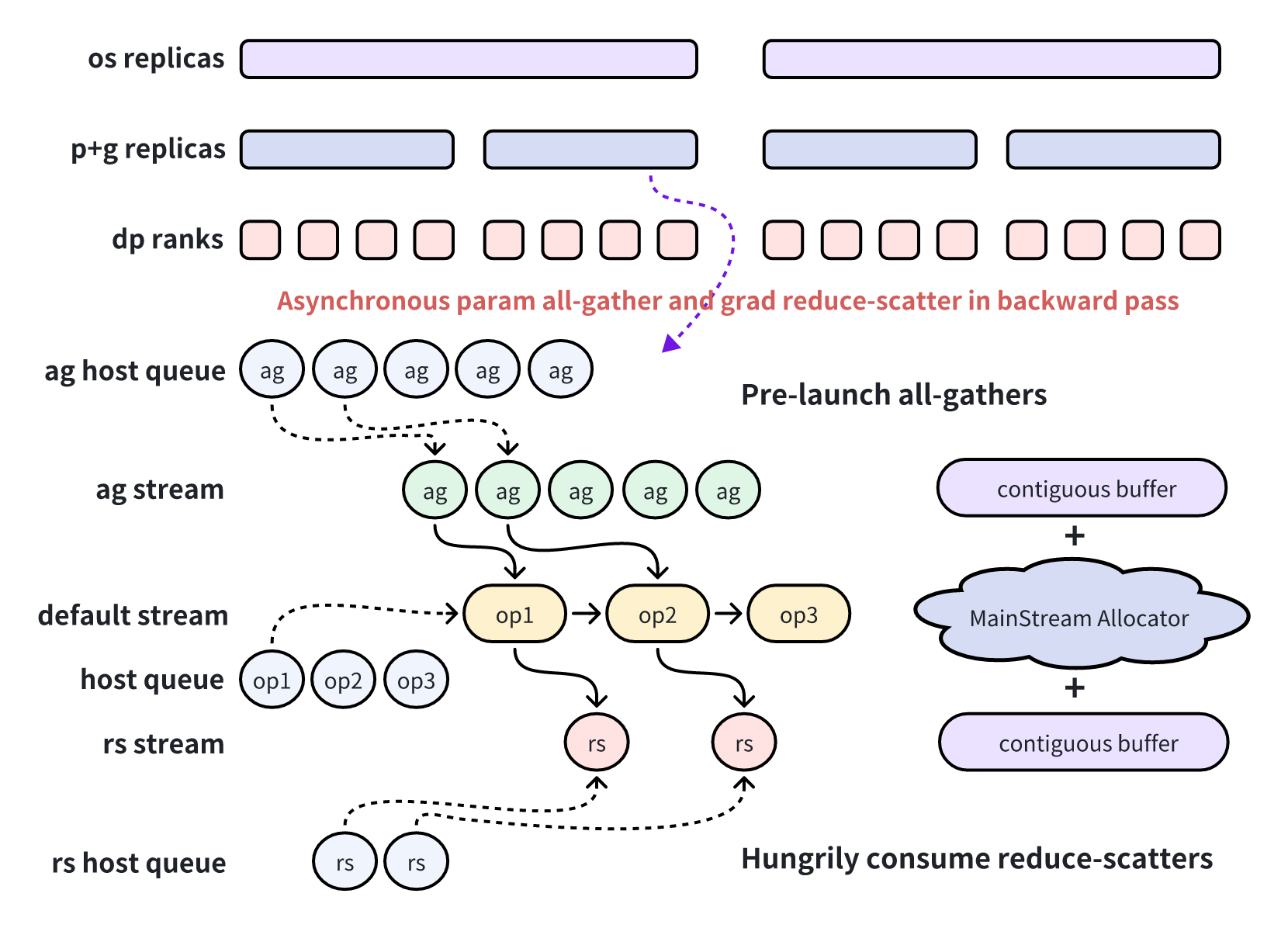}
    \caption{Illustration of Asynchronous Hierarchical Zero Parallelism. Unlike traditional Zero implementation that reshards parameters,
gradients and optimizer states across the whole Data Parallelism
dimension, AsyncHZP could adaptively reshards these with some
additional dimension, such as ZeRO1 Parallelism and ZeRO3 Parallelism. And we could asynchronously pre-launch the all-gather of parameters according to the available
memory to achieve less blocking for the layer-wise computation
and hungrily consume the reduce-scatter of gradients to reduce
memory peak.}    \label{fig:intro_plot}
\end{figure}

The classic 3D parallelism integrates data parallelism \cite{ben2019demystifying,dean2012large}, pipeline parallelism \cite{narayanan2021memory,harlap2018pipedream,huang2019gpipe}, and tensor parallelism \cite{shoeybi2019megatron} to distribute training tasks across hundreds of heterogeneous devices.
It should be noted that ND parallelism also incorporates other parallel strategies such as sequence parallelism \cite{megatronlm3}, expert parallelism\cite{lepikhin2020gshard}, etc.
This broader set of parallel techniques expands the flexibility of workload distribution, enabling more refined adaptation to the structural nuances of large-scale models.
However, designing an optimal parallelization strategy remains a non-trivial and often cumbersome task for data scientists and AI practitioners.

In contrast, ZeRO\cite{rajbhandari2020zero, rajbhandari2021zero} is a memory-efficient optimization technique designed to enable large-scale model training by eliminating memory redundancy. Its core principle is the partitioning of model states—parameters, gradients, and optimizer states—across all devices in a data-parallel group. These states are reconstructed on-the-fly via gather-based communication during training, rather than being replicated on each device.
While ND parallelism focuses on maximizing hardware utilization through multi-dimensional partitioning, ZeRO prioritizes memory efficiency and implementation simplicity, making it more accessible for practitioners seeking to scale training with minimal complexity.

As illustrated in Figure~\ref{fig:intro_plot}, we propose a novel asynchronous ZeRO variant, named Asynchronous Hierarchical Zero Parallelism (AsyncHZP), which is designed to achieve superior performance compared to ND parallelism while retaining the simplicity and memory efficiency of ZeRO. AsyncHZP introduces several key enhancements: it adaptively reshards model states across multiple ZeRO dimensions (e.g., ZeRO-1 and ZeRO-3) to optimize memory distribution; it employs a novel asynchronous scheduling method to enable non-blocking communication with minimal memory fragmentation; it asynchronously pre-launchs parameter all-gather operations to overlap communication with computation, thereby reducing layer-wise blocking; and it accelerates gradient reduce-scatter operations to lower peak memory usage. Consequently, our method exhibits superior performance compared to conventional ND parallelism in large-scale training scenarios.

In summary, the main contributions of this work are as follows:
\begin{itemize}
    \item We present a novel parallel strategy, Hierarchical ZeRO Parallelism, which adaptively reshards parameters, gradients, and optimizer states. This approach maximizes device memory utilization and achieves better performance, and it can be seamlessly integrated with traditional ND parallelism techniques to further enhance training efficiency.
    \item We design an essentially non-fragmented asynchronous scheduling method, which enables asynchronous launching of the all-gather of parameters and reduce-scatter of gradients. Compared with sequential launching, it achieves negligible memory fragmentation and non-blocking execution by cyclically reusing contiguous buffers and launching with extra threads.
    \item Through extensive experiments, we demonstrate that the proposed AsyncHZP achieves superior performance and stability when scaling to large training configurations. Our method consistently outperforms classic ND parallelism strategies on both dense and Mixture-of-Experts (MoE) models, without requiring complex strategic tuning.
\end{itemize}

\section{Related Work}

\subsection{Hierarchical ZeRO Variants}
ZeRO (Zero Redundancy Optimizer) is a key optimization technique for distributed training of large-scale models. By partitioning optimizer states, gradients, and model parameters across multiple devices, ZeRO significantly reduces the memory footprint on each GPU, thereby enabling the training of models with a substantially larger number of parameters.
Traditional ZeRO\cite{rajbhandari2020zero} is implemented in several stages, each offering a different sharding strategy. ZeRO-1 focuses on sharding optimizer states, eliminating redundant storage on each GPU. ZeRO-2 goes further by sharding gradients, while ZeRO-3 fully partitions model parameters, gradients, and optimizer states, maximizing memory savings. ZeRO-Offload\cite{rajbhandari2021zero} transfers parts of the optimizer states and gradients to CPU or disk storage, alleviating GPU memory strain.
In contrast, ZeRO++\cite{zeroplusplus} adopts a hierarchical design for better communication efficiency of parameters all-gather, using a secondary partition with a full replica at each node, which is re-partitioned from the forward parameters of all-gather across all the machines. Besides, the hybrid sharding of PyTorch FSDP\cite{fsdp2} only regularly combined the sharding and replication across each subgroup of nodes. Our AsyncHZP thoroughly shards the model states in a hierarchical manner, and we could flexibly designate the sharding group size of each model state(optimizers, gradients, and optimizer states) according to the available device memory.

\subsection{Asynchronous Execution}
In large-scale model training, asynchronous execution has emerged as a key technique to mitigate bottlenecks caused by synchronous barriers, enabling a more efficient utilization of computational resources. In the context of LLM training, Megatron-LM \cite{megatronlm} integrated a few of asynchronous execution methods, such as asynchronous gradient all-reduces, asynchronous I/O for loading massive datasets onto devices, etc. And the RL framework AsyncFlow\cite{asyncflow} innovatively designed an asynchronous pipeline to achieve extreme overlapping between different stages.
At a more fundamental level, Flux\cite{chang2024flux} overdecomposes communication and computation operations
into much finer-grained operations and achieves a better efficiency of computation-communication overlap.
Layer-wise prefill\cite{qin2024mooncake}
aims to reduce the memory occupation cost by overlapping the transferring and dumping of KV-Cache with computation.
Similarly, AsyncHZP has proposed a new asynchronous execution method that could asynchronously launch the computation and communication operators with multi-streams. The asynchronous launching reduces layer-wise blocking for parameter consumption with pre-launched all-gathers and accelerates the release of memory occupied by gradients with hungrily consumed reduce-scatters.

\section{Background and Motivation}

In this section, we first provide a brief introduction to the variants of the Zero Redundancy Optimizer (ZeRO) and analyze its limitations, particularly when deployed on large-scale computing clusters. Subsequently, we identify the communication inefficiencies that arise when integrating tensor parallelism with context parallelism methods, such as RingAttention\cite{liu2024ringattention} and Deep-Speed Ulysess\cite{jacobs2023deepspeed}.

\subsection{ZeRO Implementations and Limitations}

Typical ZeRO implementation has a key limitation that defines the ZeRO sharding group for different model states over the entire data-parallel world size. In large-scale clusters, this leads to excessively large communication groups, resulting in over-fine-grained sharding that renders communication inefficient. Specifically, partitioning model components across all data-parallel replicas is often unnecessary and counterproductive. For instance, ZeRO-3 requires all-gather and reduce-scatter operations for parameters and gradients, respectively, during each micro-batch. The latency of these collective operations scales poorly with the size of the communication group, making them unsuitable for very large groups. Besides, the optimizer states should also not be sharded endlessly with the increasing cluster.

To alleviate this issue, we propose Hierarchical ZeRO Parallelism (HZP), a novel approach that introduces separate parallel dimensions for sharding parameters, gradients, and optimizer states. This decoupling provides the flexibility to configure sharding groups independently, based on communication overhead and available memory. As a specific example, sharding parameters and gradients exclusively within a node can leverage high-speed intra-node interconnects (e.g., NVLink\cite{nvlink}) to significantly improve communication efficiency.

\subsection{Comparison with Tensor Parallelism}

Here, we mainly discuss the communication efficiency and memory consumption of Hierarchical ZeRO Parallelism (HZP) and Tensor Parallelism (TP). First, the communication volume in TP is proportional to the sequence length (number of tokens) processed per device, whereas in HZP, it is independent of the sequence length. Consequently, with a sufficiently large number of tokens per device, HZP can effectively overlap communication with computation, potentially outperforming TP in throughput-sensitive scenarios.

Second, when combining TP with intra-node context parallelism(CP) methods like RingAttention\cite{liu2024ringattention} and Deep-Speed Ulysess\cite{jacobs2023deepspeed}, the parallelism degree is constrained by the number of devices within the node. Furthermore, extending a hybrid TP and CP strategy across nodes often leads to significant performance degradation due to the high latency of inter-node communication. In contrast, HZP is orthogonal to CP. This orthogonality means HZP's parameter sharding is not restricted by the context parallelism degree, and vice versa. Both HZP and CP can concurrently utilize high-speed intra-node bandwidth without interfering with each other's parallelism dimensions, thereby ensuring higher training efficiency in a broader range of configurations.

Third, both HZP and TP are capable of reducing static memory consumption during model training, including parameters, gradients, and optimizer states. The operation of HZP is analogous to full-parameter training as it operates without intermediate tensor partitioning. In contrast, TP can achieve a partial reduction in activation memory by leveraging sequence parallelism for layers such as normalization\cite{layernorm} and dropout. However, a key limitation of TP is the necessity of storing full-shape tensors after the all-gather operation in the forward pass. Conversely, the strategic combination of HZP with context parallelism methods enables a complete reduction of activation memory via sequence partitioning.
\section{Design and Implementation}
\label{section:hzp}

Considering the complexity of ND parallelism and inefficiency of traditional ZeROs, we introduce the novel parallelism paradigm AsyncHZP, which adopts a new hierarchical design for sharding dimensions of different model states and integrates with a memory-aware asynchronous scheduling mechanism which performs nearly 100\% overlap of computation and communication.

\subsection{Hierarchical Zero Parallelism}
\label{section:hzp_runtime}

\subsubsection{Memory Sharding Analysis}

When training with mixed precision\cite{micikevicius2017mixed} using the Adam\cite{DBLP:journals/corr/Adam} optimizer, which is a common practice in the Megatron-LM\cite{megatronlm, megatron-2021, megatronlm3} framework for large LLM training. The components and corresponding static memory consumption are shown in Table~\ref{tab:hzp_memory}.

\begin{table}[htbp]
\centering
    \begin{tabular}{>{\centering\arraybackslash}ccc}
    \toprule
    \textbf{Components} & \textbf{DType} & \textbf{Memory Consumption} \\
    \midrule
    Parameters         & BF16 & 2N \\
    Gradients          & FP32 & 4N \\
    Parameter Replica  & FP32 & 4N \\
    Momentum in Adam   & FP32 & 4N \\
    Variance in Adam   & FP32 & 4N \\
    \bottomrule
    \end{tabular}
    \caption{Memory consumption with mixed-precision training (N=number of parameters).}
    \label{tab:hzp_memory}
\end{table}

Assume data parallel size is $DP$, total number of model parameters is $N$, we denote the sharding group sizes of parameters, gradients, and optimizer states as $Z_1$, $Z_2$, $Z_3$, representatively.
Therefore, the formula analysis of static memory usage is shown as follows.

For vanilla ZeRO3,
$$M_{\text{zero3}}=\frac{18N}{DP}$$
For vanilla ZeRO3 with multiple replicas, named ZeRO Parallelism (ZP),
$$M_{\text{zp}}=\frac{18N}{Z},\space (Z=Z_1=Z_2=Z_3, Z < DP)$$

And for Hierarchical Zero Parallelism (HZP), sharding group sizes for parameters, gradients, and optimizer states can be arbitrarily specified,
$$M_{\text{hzp}}=\frac{12N}{Z_1} + \frac{4N}{Z_2} + \frac{2N}{Z_3}$$

Regular sharding for all model states could not fully utilize the device resources in most cases, and HZP could properly address this issue without memory waste and inefficient communication.

\subsubsection{Workflow Illustration}
\begin{figure*}[t]
    \centering
    \includegraphics[width=1\linewidth]{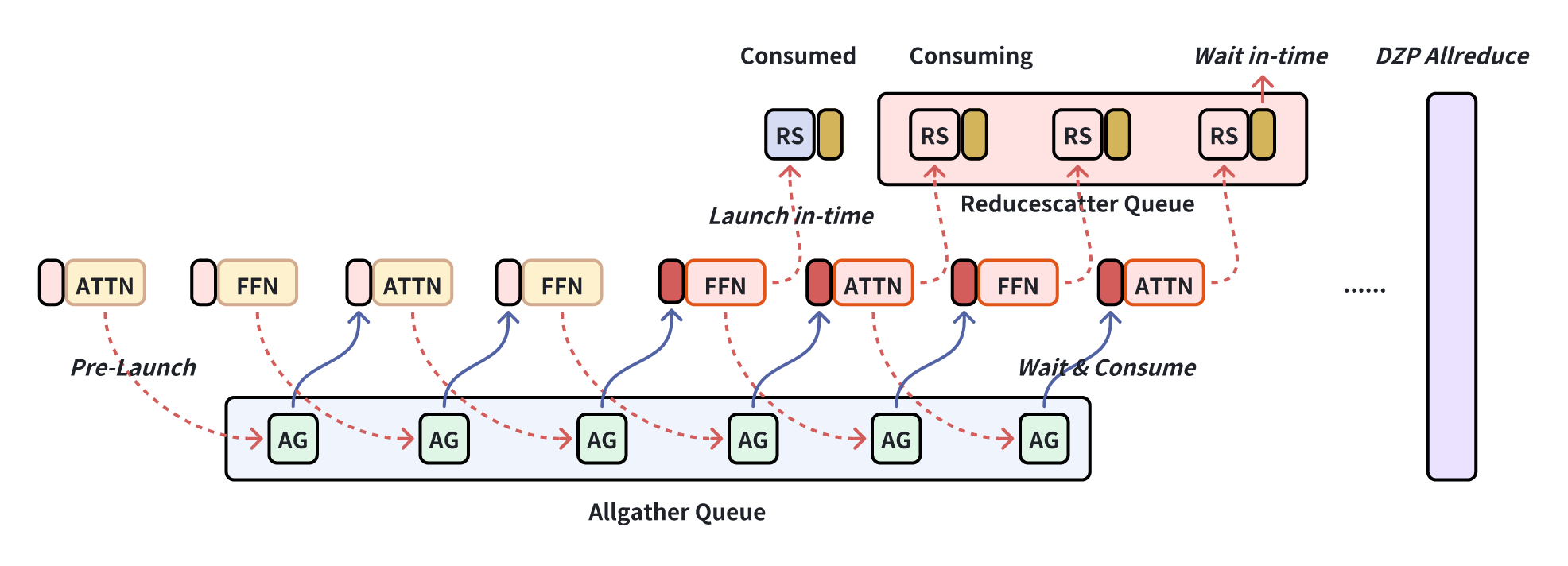}
    \caption{Illustration of Forwarding-backwarding for Hierarchical Zero Parallelism with Asynchronous Scheduling.}
    \label{fig:hzp_overall_flow_chart}
\end{figure*}

The workflow of AsyncHZP is illustrated in Figure~\ref{fig:hzp_overall_flow_chart}. ATTN and FFN denote the submodules of transformer layers for both forward and backward propagation. AG indicates all-gather of parameters during all forward and backward pass, which is pre-launched and immediately resharded after being used with no waste of memory consumption. RS indicates reduce-scatter of gradients during each micro batch's backward pass, which is launched in time and consumed subsequently for efficient memory recycling. DZP Allreduce indicates the final gradients all-reduce between different replicas of the model states. It is worth mentioning that AG and RS are executed in extra threads and will not block the execution of mainstream computing.

The implementation details can be broken down into the following steps:
\begin{itemize}
\item \textbf{Parameter Pre-fetching}: Before the execution of the first Transformer layer, one or more asynchronous all-gather operations for parameters are initiated, governed by the available device memory. This communication is designed to overlap with data loading and other preceding computations.
\item \textbf{Forward Pass}: During the forward pass, we utilize a pipelined pre-fetching mechanism. All-gather operations for the parameters of subsequent layers are continuously launched in advance, ensuring they are available just-in-time for computation and thus preventing pipeline stalls. For the transition between the final layer of the forward pass and the first layer of the backward pass, parameters are explicitly re-gathered.
\item \textbf{Backward Pass}: The backward pass similarly requires all-gather operations to reconstruct model parameters needed for gradient computation. In parallel, as gradients are computed for each layer, a reduce-scatter operation is immediately initiated to aggregate them. This operation is scheduled asynchronously to avoid blocking other computations. To ensure numerical precision alignment with established frameworks like Megatron-LM\cite{megatronlm, megatron-2021, megatronlm3}, all reduce-scatter operations are performed using the float32 data type.
\item \textbf{Optimizer Step}: After all micro-batches are processed, a final all-reduce operation is performed across the data-parallel group to aggregate the gradients from all replicas. The optimizer then updates its local state shards (ZeRO-1) using these aggregated gradient slices. Subsequently, an asynchronous all-gather is executed. This operation effectively reconstructs the updated model parameter shards (ZeRO-3) from the optimizer states, preparing them for the next training iteration.

\end{itemize}

\begin{figure*}[t]
    \centering
    \includegraphics[width=1\linewidth]{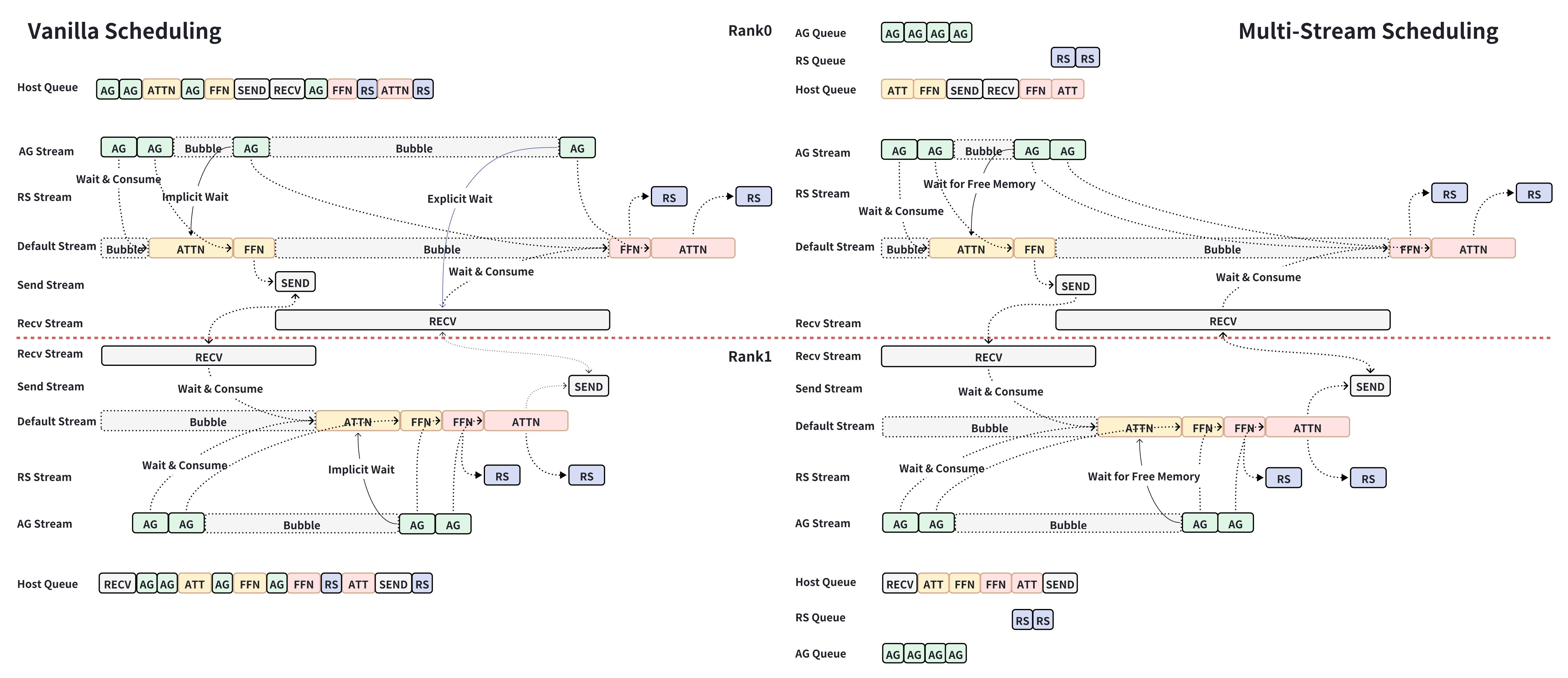}
    \caption{Multi-stream asynchronous scheduling for non-blocking communications.}
    \label{fig:pipeline_async_schedule}
\end{figure*}

\subsection{Asynchronous Scheduling}

Hierarchical ZeRO Parallelism partitions parameters, gradients, and optimizer states, and inserts necessary communication to achieve equivalent computation as full-parameter training. It is vital to better overlap communication with computation to achieve higher Model FLOPS Utilization. All-gather is invoked before forward and backward execution to get full parameters for subsequent computation. And reduce-scatter is invoked to aggregate gradients of every micro batch for gradient partitions.
The communication for each weight or gradient has temporal dependency with the corresponding computation kernel, but is independent of each other. One approach for overlap is to group weights and gradients and partition them into multiple buckets, and communication is parallel with the computation of different bucket. However, in scenarios involving gradient accumulation and pipeline parallelism, each parameter undergoes multiple aggregations, and the computation graph becomes more complex. It is more flexible and effective to launch communication based on parameter granularity.

As illustrated in Figure~\ref{fig:pipeline_async_schedule}, a vanilla method for asynchronous scheduling is leveraging the previously recorded schedules to pre-launch an all-gather for the next module's parameters and manually waiting the sequentially launched reduce-scatters, as shown on the left. However, the all-gather and reduce-scatter operations would insert a synchronous event at the issued point on both the communication stream and the default stream to enforce correct inter-stream ordering, resulting in a worse overlap between communication and computation. To address this issue, we proposed a multi-stream asynchronous scheduling method with asynchronous dispatch as shown on the right. Together with a specifically designed memory allocator, we could pre-allocate a fixed-capacity, persistent memory pool dedicated to asynchronous commands, including all-gathers and reduce-scatters. All-gather of parameters could be asynchronously pre-launched to meet the computational consumption without any waiting bubbles, and the reduce-scatter would not insert a synchronous event, ensuring the completion of communication and the release of occupied device memory, which could be quickly consumed in an extra thread.
Compared to sequentially launching all the operations with the main thread, the reuse between computation and communication streams of device memory is not enabled with our method, but the contiguous sub-stream buffers, reused cyclically by a single type of operation, allow for negligible memory fragmentation.

\subsection{Integrated with ND Parallelism}
\label{section:hzp_scalability}
Our AsyncHZP is lightweight and user-friendly, enabling deep integration and optimization with other native parallelism and optimization features in Megatron-LM. This significantly improves model training efficiency and fully leverages the advantages of other Megatron ND parallelism techniques. In this section, we will mainly introduce some strategies that have special optimizations and advantages coupled with AsyncHZP.

\subsubsection{AsyncHZP with Context Parallelism}

\begin{figure}[t]
    \centering
    \includegraphics[width=1\linewidth]{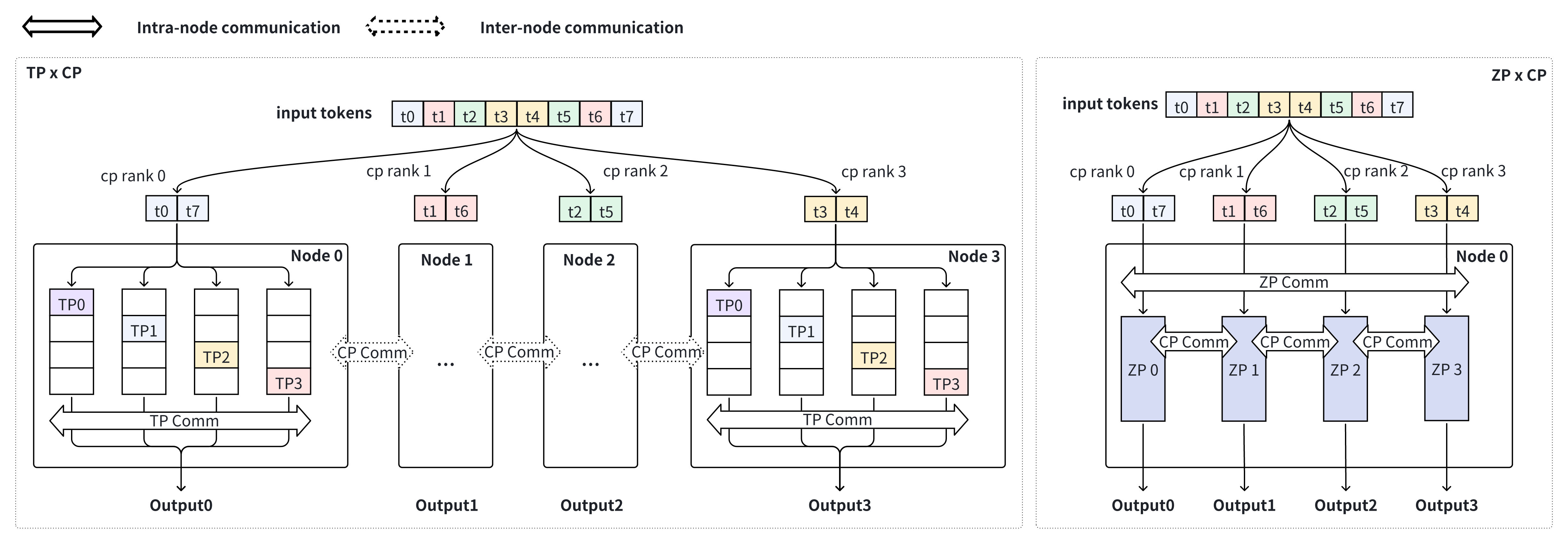}
    \caption{Comparison between tensor parallelism(TP) and AsyncHZP when coupled with context parallelism(CP).}
    \label{fig:hzpwithcp}
\end{figure}

Context parallelism involves splitting the samples across multiple devices, which could significantly reduce the activation memory consumption, especially for long sequences.
Common implementation strategies for context parallelism include DeepSpeed-Ulysses\cite{jacobs2023deepspeedulyssesoptimizationsenabling}, RingAttention\cite{liu2024ringattention}, etc. As shown in Figure~\ref{fig:hzpwithcp}, our AsyncHZP could share communication groups with context parallelism, compared to Tensor Parallelism (TP). We could effectively reduce the static memory used by the model states as TP, and we could also avoid inefficient cross-machine communications when training long sequences with context parallelism.

\subsubsection{AsyncHZP with Pipeline Parallelism}
\label{section:hzp_zp_pp}

\begin{figure}[t]
    \centering
    \includegraphics[width=1\linewidth]{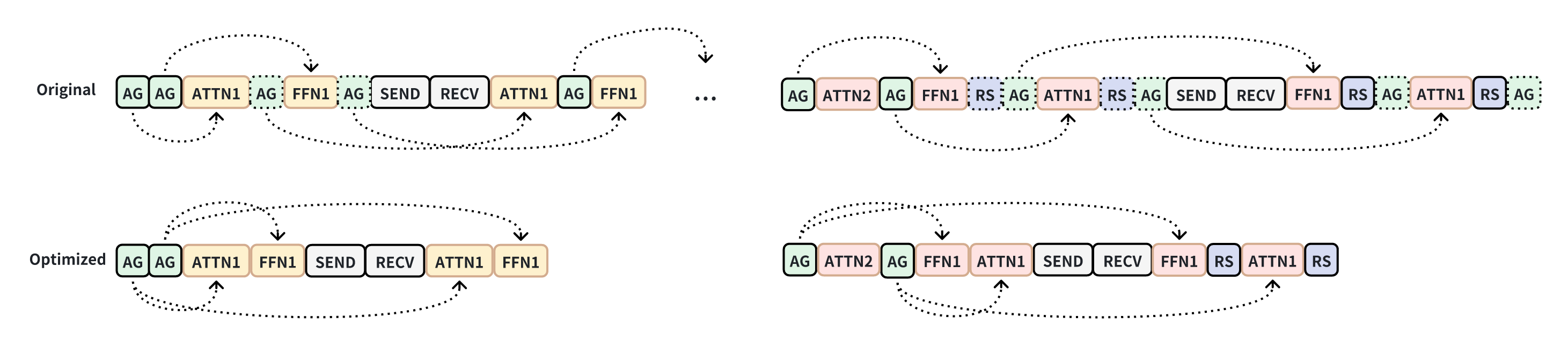}
    \caption{AsyncHZP with Pipeline Parallelism with shared communication operations.}
    \label{fig:hzpwithpp}
\end{figure}

The advantage of PP lies in its multiple variants (e.g., interleaved1F1B\cite{narayanan2021efficient}, ZeroBubble\cite{qi2023zero}, Terapipe\cite{li2021terapipe}), which offer flexibility for diverse task scenarios. Combining AsyncHZP with pipeline parallelism, there exist some optimizations that could further reduce the communication cost and achieve efficient communication overlap. In interleaved1F1B,
during the warm-up and cool-down phase of pipeline parallelism, the adjacent micro batches within the same virtual pipeline stage could reuse the corresponding all-gather of parameters. The reduce-scatter of gradients could also be reused with accumulated gradients for the adjacent micro batches during the cool-down phase. The 1F1B steady-state schedule facilitates a specific form of parameter reuse across non-adjacent, like-kind passes. The forward pass for micro-batch $k+1$ ($F_{k+1}$) can reuse the parameters from the preceding forward pass ($F_k$) by simply caching them across one intervening backward pass.
As illustrated in Figure~\ref{fig:hzpwithpp}, the optimized execution flow of AsyncHZP with pipeline parallelism is more efficient.

\subsubsection{AsyncHZP with Recomputation}

In terms of performance, enabling full recomputation with AsyncHZP incurs no additional communication overhead compared to disabling full recomputation.
Since AsyncHZP requires a parameter all-gather before the backward pass, it eliminates the need for extra parameter gathering operations for recomputation of the forward pass.
In contrast, the same communication operations are necessary for TP during recomputation as in the forward pass, resulting in higher recomputation costs.
Furthermore, there is no intrusive modification of the training logic except for simple hooks for the forward and backward pass of modules.

\subsection{Scalability Analysis}

Within a regular transformer layer, Tensor parallelism introduces 8x communication operations, including 4x all-gathers and 4x reduce-scatters, whose volumes of communication are positively correlated with the sequence length. In contrast, the communication volume of AsyncHZP is invariant to the context length, being solely determined by the static shape of the weight tensors. In long-sequence or large micro-batch size scenarios, while TP requires relatively higher communication costs that are proportional to context length, AsyncHZP circumvents this overhead entirely and exhibits strong competitiveness.
Furthermore, AsyncHZP is capable of enabling extreme partitioning of optimizer parameters, allowing smaller sharding groups for parameters and gradients that do not block each other during forward and backward propagation of different accumulation steps.
Thus, we can significantly reduce the load imbalance of different data parallelism ranks with length-varying samples, which requires uneven computational costs for self-attention modules.
To conclude, our AsyncHZP is scalable for long sequences and massive clusters training with larger Data Parallelism group, on account of its hierarchical design and compatibility with context parallelism.

\begin{table}[h]
\centering
\begin{threeparttable}
\begin{tabularx}{\linewidth}{c>{\centering\arraybackslash}Xccc}
\toprule
\textbf{Model} & \textbf{Parallel Strategies} &
\textbf{MFU} & \textbf{Memory (GB)} & \textbf{SpeedUp} \\
\midrule
\multirow{2}{*}{Seed-OSS-9B-Base}
& $Z_1$=64, $Z_2$=$Z_3$=8, $CP$=8 & 40.41\% & 61.57 & 119.5\% \\
& $TP$=8 & 33.83\% & 52.58 & 100.0\% \\
\midrule
\multirow{2}{*}{Seed-OSS-36B-Base}
& $Z_1$=64, $Z_2$=$Z_3$=8, $PP$=4, $CP$=8 & 46.26\% & 57.98 & 113.2\% \\
& $TP$=8, $PP$=4 & 40.85\% & 52.34 & 100.0\% \\
\midrule
\multirow{2}{*}{MoE-100B}
& $Z_1$=64, $Z_2$=$Z_3$=8, $PP$=6, $CP$=8 & 27.10\% & 52.70 & 110.2\% \\
& $EP$=8, $PP$=6, $CP$=8 & 24.58\% & 57.74 & 100.0\% \\
\bottomrule
\end{tabularx}
\caption{Comparisons between AsyncHZP and ND parallelism with optimal parallel strategies. The total cards used for Seed-OSS and MoE models are separately 256 and 384.}
\label{tab:evaluation_parallel_strategy_models_256gpus}
\end{threeparttable}
\end{table}

\section{Experiments}

\subsection{Experiment Setup}

All experiments are conducted on Megatron-LM\cite{megatronlm, megatron-2021, megatronlm3} for convenient comparison and integration with traditional ND parallelism techniques, which is a popular choice for training large-scale models. Our AsyncHZP is developed incrementally on the basis of this framework. We evaluate our method on both Dense models(Seed-Oss\cite{seed2025seed-oss}) and DeepseekV3-like \cite{deepseekai2024deepseekv3technicalreport, fast-moe, google_glam} Mixture-of-Experts models with maximum sequence length 32K for packed samples, global batch 16M tokens and default 256 cards. The metrics we adopt to analyze the performance of different training configurations are Model FLOPs Utilization(MFU), memory usage, and cluster linearity. Additionally, it should be noted that recomputation and activations offloading are used as needed to reduce memory usage, which is not discussed in the following experiments.

The sheer scale of these experiments, characterized by massive model sizes and extensive data processing, necessitates access to significant computational resources. The prevailing paradigm for training large-scale language models has historically been reliant on advanced Graphics Processing Units (GPUs), largely attributable to their mature software ecosystems and vibrant developer communities. This GPU-centric approach is evidenced by the development of state-of-the-art open-source models, including LLaMA-3\cite{dubey2024llama}, DeepSeek-R1\cite{guo2025deepseek}, and Qwen series\cite{qwen2, qwen2.5, qwen3}, which were all trained from scratch on GPU-based infrastructures. However, the acquisition of large-scale training clusters has become exceptionally challenging due to persistent global supply chain constraints. A pivotal recent development challenging this status quo is that Google has announced the successful end-to-end training of its Gemini model\cite{team2024gemini} utilizing its latest-generation, in-house developed Tensor Processing Units (TPUs)\cite{jouppi2023tpu}, thereby challenging the long-standing training paradigm dominated by GPUs.

In response to these industry-wide challenges and inspired by this trend, our work adopts an alternative AI accelerator, each card providing 300+ TFLOPS of BF16 computing power and 64 GB of high-bandwidth memory (HBM). The accelerators are deployed in a configuration of 16 units per node and each group with 8 accelerators are interconnected via a full-mesh topology to facilitate high-speed communication. Besides, each card is equipped with high-speed, NVLink-like interconnects, which collectively form a full-mesh topology to deliver an aggregate intra-node bandwidth of approximately 400 GB/s, while inter-node connectivity is achieved using RDMA over Converged Ethernet (RoCE), which provides an interconnect bandwidth of 200 Gbps.

\subsection{Comparison with ND parallelism}
\label{section:evaluation}

In this section, we compare our AsyncHZP with classic ND parallelism techniques, and all the models, from Dense to Mix-of-Experts, are tuned to optimal performance. As shown in Table~\ref{tab:evaluation_parallel_strategy_models_256gpus}, our AsyncHZP achieves about 25\% higher performance than ND parallelism strategies. As for detailed analysis, we could find that the device memory could be fully utilized while training with larger models, and the sharding groups for parameters and gradients often are not distributed across machines for better communication efficiency.


\subsection{Ablation study on Asynchronous Scheduling}

For better overlap between communication and computation, we propose an asynchronous scheduling method for non-blocking communications. These comparisons (Table~\ref{tab:evaluation_parallel_strategy_models_256gpus_async}) clearly demonstrate the effectiveness of our method, achieving comparable performance or above.
Compared to sequentially launching all the operations, we could asynchronously launch all the all-gathers of parameters and reduce-scatters of gradients without blocking.
We could better overlap the communication costs with computation and accelerate the release of occupied device memory.
\begin{table}[h]
\centering
\begin{threeparttable}
\begin{tabularx}{\linewidth}{c>{\centering\arraybackslash}Xccc}
\toprule
\textbf{Model} & \textbf{Parallel Strategies} & \textbf{Async Scheduling} &
\textbf{MFU} & \textbf{SpeedUp} \\
\midrule
\multirow{2}{*}{Seed-OSS-36B-Base-8K} & \multirow{2}{*}{$Z_1$=64, $Z_2$=$Z_3$=8, $PP$=8}
& $\checkmark$ & 44.40\% & 102.9\% \\
& & $\times$ & 43.12\% & 100.0\% \\
\midrule

\multirow{2}{*}{Seed-OSS-36B-Base-8K} & \multirow{2}{*}{$Z_1$=256, $Z_2$=$Z_3$=8, $CP$=8}
& $\checkmark$ & 49.50\% & 100.3\% \\
& & $\times$ & 49.32\% & 100.0\% \\
\midrule
\bottomrule
\end{tabularx}
\caption{Performance improvements for asynchronous scheduling. $\times$ indicates training with the vanilla scheduling method. }
\label{tab:evaluation_parallel_strategy_models_256gpus_async}
\end{threeparttable}
\end{table}

\subsection{Ablation study on Context Length}

In addition to the efficient combination of AsyncHZP and context parallelism for long-sequence training, our method also shows a strong generalization for short-sequence training without obvious performance degradation.
As shown in Table~\ref{tab:hzp_context_length}, we could achieve comparable performance in short-length cases and great improvements in longer-length cases.

\begin{table}[h]
\centering
\begin{threeparttable}
\begin{tabularx}{\linewidth}{>{\centering\arraybackslash}cXccc}
\toprule
\textbf{Context Length} & \textbf{Parallel Strategies}& \textbf{MFU} & \textbf{Memory (GB)} & \textbf{SpeedUp}\\
\midrule
\multirow{2}{*}{8K}
  & $Z_1$=256, $Z_2$=$Z_3$=8, $CP$=8      & 49.50\% & 63.87  & 115.2\% \\
  & $TP$=8                        & 42.98\%  & 58.46 & 100.0\% \\
\midrule
\multirow{2}{*}{32K}
  & $Z_1$=64, $Z_2$=$Z_3$=8, $PP$=4, $CP$=8 & 46.26\% & 57.98 & 113.2\% \\
  & $TP$=8, $PP$=4 & 40.85\% & 52.34 & 100.0\% \\
\midrule
\multirow{2}{*}{128K}
  & $Z_1$=64, $Z_2$=$Z_3$=8, $PP$=4, $CP$=8     & 33.64\% & 59.42 & 116.7\% \\
  & $TP$=8, $PP$=4, $CP$=8                        & 28.82\% & 62.42 & 100.0\% \\
\bottomrule
\end{tabularx}
\caption{Comparisons between AsyncHZP and ND parallelism with different context length on Seed-OSS 36B.}
\label{tab:hzp_context_length}
\end{threeparttable}
\end{table}

\subsection{Ablation study on Cluster scalability}

During training with large-scale clusters, we inevitably encounter performance degradation in most scenarios. Thus, ensuring extremely high scalability is a significant challenge for us to train large models in massive clusters. Owing to the hierarchical design of AsyncHZP, we could achieve better scalability by enabling more intra-machine communication. The results of the scalability evaluation are shown in Table~\ref{tab:evaluation_parallel_strategy_36b}. In contrast, the cluster linearity of optimal ND parallelism strategies is 88.37\%, versus 91.12\% for AsyncHZP, while scaling from 256 cards to 1024 cards. The sharp degradation of cluster linearity for both methods is mainly caused by the variable-length input sequences with uneven computational costs for self-attention modules.

\begin{table}[h]
\centering
\begin{threeparttable}
\begin{tabularx}{\linewidth}{>{\centering\arraybackslash}cXccc}
\toprule
\textbf{Total Cards} & \textbf{Parallel Strategies} & \textbf{MFU} & \textbf{Linearity} \\
\midrule
\multirow{1}{*}{256}
 & $Z_1$=64, $Z_2$=$Z_3$=8, $PP$=4, $CP$=8 & 46.26\%  & 100.0\% \\
\midrule
\multirow{1}{*}{512}
 & $Z_1$=64, $Z_2$=$Z_3$=8, $PP$=4, $CP$=8  & 45.57\% & 98.51\% \\
\midrule
\multirow{1}{*}{1024}
 & $Z_1$=64, $Z_2$=$Z_3$=8, $PP$=4, $CP$=8  & 42.15\% & 91.12\% \\
\bottomrule
\end{tabularx}
\caption{The results of linear scalability on Seed-OSS 36B. Our AsyncHZP could achieve better linearity while scaling to larger clusters.}
\label{tab:evaluation_parallel_strategy_36b}
\end{threeparttable}
\end{table}

\section{Conclusion}

This paper presents Asynchronous Hierarchical Zero Parallelism (AsyncHZP), a novel and efficient ZeRO variant designed to enhance memory utilization and performance scalability in large-scale training scenarios. Our approach makes two primary contributions. First, it employs an adaptive hierarchical resharding strategy for parameters, gradients, and optimizer states, which optimizes device memory footprint and substantially reduces communication overhead. Second, we introduce a multi-stream asynchronous scheduling mechanism that asynchronously prefetches model parameters and completes gradient reduction with extra threads. This design maximizes the overlap between computation and communication with minimal overhead. Extensive experiments on both Dense and Mixture-of-Experts (MoE) models validate that AsyncHZP achieves superior performance and robust scalability, outperforming classic ND parallelism without requiring complex strategic tuning. Furthermore, our method maintains seamless compatibility with existing ND parallelism techniques, offering a powerful yet simple solution for state-of-the-art large-scale training.

\clearpage

\bibliographystyle{plainnat}
\bibliography{main}

\begin{thebibliography}{40}
\providecommand{\natexlab}[1]{#1}
\providecommand{\url}[1]{\texttt{#1}}
\expandafter\ifx\csname urlstyle\endcsname\relax
  \providecommand{\doi}[1]{doi: #1}\else
  \providecommand{\doi}{doi: \begingroup \urlstyle{rm}\Url}\fi

\bibitem[Ben-Nun and Hoefler(2019)]{ben2019demystifying}
Tal Ben-Nun and Torsten Hoefler.
\newblock Demystifying parallel and distributed deep learning: An in-depth concurrency analysis.
\newblock \emph{ACM Computing Surveys (CSUR)}, 52\penalty0 (4):\penalty0 1--43, 2019.

\bibitem[Chang et~al.(2024)Chang, Bao, Hou, Jiang, Zheng, Zhong, Zhang, Song, Jiang, Lin, Jin, and Liu]{chang2024flux}
Li-Wen Chang, Wenlei Bao, Qi~Hou, Chengquan Jiang, Ningxin Zheng, Yinmin Zhong, Xuanrun Zhang, Zuquan Song, Ziheng Jiang, Haibin Lin, Xin Jin, and Xin Liu.
\newblock Flux: Fast software-based communication overlap on gpus through kernel fusion, 2024.

\bibitem[Dean et~al.(2012)Dean, Corrado, Monga, Chen, Devin, Mao, Ranzato, Senior, Tucker, Yang, et~al.]{dean2012large}
Jeffrey Dean, Greg Corrado, Rajat Monga, Kai Chen, Matthieu Devin, Mark Mao, Marc'aurelio Ranzato, Andrew Senior, Paul Tucker, Ke~Yang, et~al.
\newblock Large scale distributed deep networks.
\newblock \emph{Advances in neural information processing systems}, 25, 2012.

\bibitem[DeepSeek-AI(2024)]{deepseekai2024deepseekv3technicalreport}
DeepSeek-AI.
\newblock Deepseek-v3 technical report, 2024.
\newblock URL \url{https://arxiv.org/abs/2412.19437}.

\bibitem[Dehghani et~al.(2021)Dehghani, Arnab, Beyer, Vaswani, and Tay]{layernorm}
Mostafa Dehghani, Anurag Arnab, Lucas Beyer, Ashish Vaswani, and Yi~Tay.
\newblock The efficiency misnomer.
\newblock \emph{ArXiv}, abs/2110.12894, 2021.

\bibitem[Du et~al.(2021)Du, Huang, Dai, Tong, Lepikhin, Xu, Krikun, Zhou, Yu, Firat, et~al.]{google_glam}
Nan Du, Yanping Huang, Andrew~M Dai, Simon Tong, Dmitry Lepikhin, Yuanzhong Xu, Maxim Krikun, Yanqi Zhou, Adams~Wei Yu, Orhan Firat, et~al.
\newblock Glam: Efficient scaling of language models with mixture-of-experts.
\newblock \emph{arXiv preprint arXiv:2112.06905}, 2021.

\bibitem[Dubey et~al.(2024)Dubey, Jauhri, Pandey, Kadian, Al-Dahle, Letman, Mathur, Schelten, Yang, Fan, et~al.]{dubey2024llama}
Abhimanyu Dubey, Abhinav Jauhri, Abhinav Pandey, Abhishek Kadian, Ahmad Al-Dahle, Aiesha Letman, Akhil Mathur, Alan Schelten, Amy Yang, Angela Fan, et~al.
\newblock The llama 3 herd of models.
\newblock \emph{arXiv e-prints}, pages arXiv--2407, 2024.

\bibitem[Guo et~al.(2025)Guo, Yang, Zhang, Song, Zhang, Xu, Zhu, Ma, Wang, Bi, et~al.]{guo2025deepseek}
Daya Guo, Dejian Yang, Haowei Zhang, Junxiao Song, Ruoyu Zhang, Runxin Xu, Qihao Zhu, Shirong Ma, Peiyi Wang, Xiao Bi, et~al.
\newblock Deepseek-r1: Incentivizing reasoning capability in llms via reinforcement learning.
\newblock \emph{arXiv preprint arXiv:2501.12948}, 2025.

\bibitem[Han et~al.(2025)Han, You, Wang, Luo, Yang, Shi, Chen, Zhang, Lan, Deng, Ji, Liu, Huang, Zhang, Pan, Wang, Huang, Li, and Wu]{asyncflow}
Zhenyu Han, Ansheng You, Haibo Wang, Kui Luo, Guang Yang, Wenqi Shi, Menglong Chen, Sicheng Zhang, Zeshun Lan, Chunshi Deng, Huazhong Ji, Wenjie Liu, Yu~Huang, Yixiang Zhang, Chenyi Pan, Jing Wang, Xin Huang, Chunsheng Li, and Jianping Wu.
\newblock Asyncflow: An asynchronous streaming rl framework for efficient llm post-training, 2025.
\newblock URL \url{https://arxiv.org/abs/2507.01663}.

\bibitem[Harlap et~al.(2018)Harlap, Narayanan, Phanishayee, Seshadri, Devanur, Ganger, and Gibbons]{harlap2018pipedream}
Aaron Harlap, Deepak Narayanan, Amar Phanishayee, Vivek Seshadri, Nikhil Devanur, Greg Ganger, and Phil Gibbons.
\newblock Pipedream: Fast and efficient pipeline parallel dnn training.
\newblock \emph{arXiv preprint arXiv:1806.03377}, 2018.

\bibitem[He et~al.(2021)He, Qiu, Zeng, Yang, Zhai, and Tang]{fast-moe}
Jiaao He, Jiezhong Qiu, Aohan Zeng, Zhilin Yang, Jidong Zhai, and Jie Tang.
\newblock Fastmoe: {A} fast mixture-of-expert training system.
\newblock \emph{CoRR}, abs/2103.13262, 2021.
\newblock URL \url{https://arxiv.org/abs/2103.13262}.

\bibitem[Huang et~al.(2019)Huang, Cheng, Bapna, Firat, Chen, Chen, Lee, Ngiam, Le, Wu, et~al.]{huang2019gpipe}
Yanping Huang, Youlong Cheng, Ankur Bapna, Orhan Firat, Dehao Chen, Mia Chen, HyoukJoong Lee, Jiquan Ngiam, Quoc~V Le, Yonghui Wu, et~al.
\newblock Gpipe: Efficient training of giant neural networks using pipeline parallelism.
\newblock \emph{Advances in neural information processing systems}, 32, 2019.

\bibitem[Jacobs et~al.(2023{\natexlab{a}})Jacobs, Tanaka, Zhang, Zhang, Song, Rajbhandari, and He]{jacobs2023deepspeed}
Sam~Ade Jacobs, Masahiro Tanaka, Chengming Zhang, Minjia Zhang, Shuaiwen~Leon Song, Samyam Rajbhandari, and Yuxiong He.
\newblock Deepspeed ulysses: System optimizations for enabling training of extreme long sequence transformer models.
\newblock \emph{arXiv preprint arXiv:2309.14509}, 2023{\natexlab{a}}.

\bibitem[Jacobs et~al.(2023{\natexlab{b}})Jacobs, Tanaka, Zhang, Zhang, Song, Rajbhandari, and He]{jacobs2023deepspeedulyssesoptimizationsenabling}
Sam~Ade Jacobs, Masahiro Tanaka, Chengming Zhang, Minjia Zhang, Shuaiwen~Leon Song, Samyam Rajbhandari, and Yuxiong He.
\newblock Deepspeed ulysses: System optimizations for enabling training of extreme long sequence transformer models, 2023{\natexlab{b}}.
\newblock URL \url{https://arxiv.org/abs/2309.14509}.

\bibitem[Jouppi et~al.(2023)Jouppi, Kurian, Li, Ma, Nagarajan, Nai, Patil, Subramanian, Swing, Towles, et~al.]{jouppi2023tpu}
Norm Jouppi, George Kurian, Sheng Li, Peter Ma, Rahul Nagarajan, Lifeng Nai, Nishant Patil, Suvinay Subramanian, Andy Swing, Brian Towles, et~al.
\newblock Tpu v4: An optically reconfigurable supercomputer for machine learning with hardware support for embeddings.
\newblock In \emph{Proceedings of the 50th annual international symposium on computer architecture}, pages 1--14, 2023.

\bibitem[Kingma and Ba(2015)]{DBLP:journals/corr/Adam}
Diederik~P. Kingma and Jimmy Ba.
\newblock Adam: {A} method for stochastic optimization.
\newblock In Yoshua Bengio and Yann LeCun, editors, \emph{3rd International Conference on Learning Representations, {ICLR} 2015, San Diego, CA, USA, May 7-9, 2015, Conference Track Proceedings}, 2015.
\newblock URL \url{http://arxiv.org/abs/1412.6980}.

\bibitem[Korthikanti et~al.(2023)Korthikanti, Casper, Lym, McAfee, Andersch, Shoeybi, and Catanzaro]{megatronlm3}
Vijay~Anand Korthikanti, Jared Casper, Sangkug Lym, Lawrence McAfee, Michael Andersch, Mohammad Shoeybi, and Bryan Catanzaro.
\newblock Reducing activation recomputation in large transformer models.
\newblock In \emph{MLSys}, 2023.
\newblock URL \url{https://proceedings.mlsys.org/paper_files/paper/2023/hash/80083951326cf5b35e5100260d64ed81-Abstract-mlsys2023.html}.

\bibitem[Lepikhin et~al.(2020)Lepikhin, Lee, Xu, Chen, Firat, Huang, Krikun, Shazeer, and Chen]{lepikhin2020gshard}
Dmitry Lepikhin, HyoukJoong Lee, Yuanzhong Xu, Dehao Chen, Orhan Firat, Yanping Huang, Maxim Krikun, Noam Shazeer, and Zhifeng Chen.
\newblock Gshard: Scaling giant models with conditional computation and automatic sharding.
\newblock \emph{arXiv preprint arXiv:2006.16668}, 2020.

\bibitem[Li et~al.(2021)Li, Zhuang, Guo, Zhuo, Zhang, Song, and Stoica]{li2021terapipe}
Zhuohan Li, Siyuan Zhuang, Shiyuan Guo, Danyang Zhuo, Hao Zhang, Dawn Song, and Ion Stoica.
\newblock Terapipe: Token-level pipeline parallelism for training large-scale language models.
\newblock In \emph{International Conference on Machine Learning}, pages 6543--6552. PMLR, 2021.

\bibitem[Liu et~al.(2024)Liu, Zaharia, and Abbeel]{liu2024ringattention}
Hao Liu, Matei Zaharia, and Pieter Abbeel.
\newblock Ringattention with blockwise transformers for near-infinite context.
\newblock In \emph{The Twelfth International Conference on Learning Representations}, 2024.
\newblock URL \url{https://openreview.net/forum?id=WsRHpHH4s0}.

\bibitem[Micikevicius et~al.(2017)Micikevicius, Narang, Alben, Diamos, Elsen, Garcia, Ginsburg, Houston, Kuchaiev, Venkatesh, et~al.]{micikevicius2017mixed}
Paulius Micikevicius, Sharan Narang, Jonah Alben, Gregory Diamos, Erich Elsen, David Garcia, Boris Ginsburg, Michael Houston, Oleksii Kuchaiev, Ganesh Venkatesh, et~al.
\newblock Mixed precision training.
\newblock \emph{arXiv preprint arXiv:1710.03740}, 2017.

\bibitem[Narayanan et~al.(2021{\natexlab{a}})Narayanan, Phanishayee, Shi, Chen, and Zaharia]{narayanan2021memory}
Deepak Narayanan, Amar Phanishayee, Kaiyu Shi, Xie Chen, and Matei Zaharia.
\newblock Memory-efficient pipeline-parallel dnn training.
\newblock In \emph{International Conference on Machine Learning}, pages 7937--7947. PMLR, 2021{\natexlab{a}}.

\bibitem[Narayanan et~al.(2021{\natexlab{b}})Narayanan, Shoeybi, Casper, LeGresley, Patwary, Korthikanti, Vainbrand, Kashinkunti, Bernauer, Catanzaro, Phanishayee, and Zaharia]{megatron-2021}
Deepak Narayanan, Mohammad Shoeybi, Jared Casper, Patrick LeGresley, Mostofa Patwary, Vijay Korthikanti, Dmitri Vainbrand, Prethvi Kashinkunti, Julie Bernauer, Bryan Catanzaro, Amar Phanishayee, and Matei Zaharia.
\newblock Efficient large-scale language model training on gpu clusters using megatron-lm.
\newblock In \emph{Proceedings of the International Conference for High Performance Computing, Networking, Storage and Analysis}, SC '21, New York, NY, USA, 2021{\natexlab{b}}. Association for Computing Machinery.
\newblock ISBN 9781450384421.
\newblock \doi{10.1145/3458817.3476209}.
\newblock URL \url{https://doi.org/10.1145/3458817.3476209}.

\bibitem[Narayanan et~al.(2021{\natexlab{c}})Narayanan, Shoeybi, Casper, LeGresley, Patwary, Korthikanti, Vainbrand, Kashinkunti, Bernauer, Catanzaro, et~al.]{narayanan2021efficient}
Deepak Narayanan, Mohammad Shoeybi, Jared Casper, Patrick LeGresley, Mostofa Patwary, Vijay Korthikanti, Dmitri Vainbrand, Prethvi Kashinkunti, Julie Bernauer, Bryan Catanzaro, et~al.
\newblock Efficient large-scale language model training on gpu clusters using megatron-lm.
\newblock In \emph{Proceedings of the international conference for high performance computing, networking, storage and analysis}, pages 1--15, 2021{\natexlab{c}}.

\bibitem[NVLink()]{nvlink}
NVLink.
\newblock {NVIDIA NVLINK}.
\newblock \url{http://www.nvidia.com/object/nvlink.html}, 2017.

\bibitem[Qi et~al.(2023)Qi, Wan, Huang, and Lin]{qi2023zero}
Penghui Qi, Xinyi Wan, Guangxing Huang, and Min Lin.
\newblock Zero bubble pipeline parallelism.
\newblock \emph{arXiv preprint arXiv:2401.10241}, 2023.

\bibitem[Qin et~al.(2024)Qin, Li, He, Zhang, Wu, Zheng, and Xu]{qin2024mooncake}
Ruoyu Qin, Zheming Li, Weiran He, Mingxing Zhang, Yongwei Wu, Weimin Zheng, and Xinran Xu.
\newblock Mooncake: A kvcache-centric disaggregated architecture for llm serving.
\newblock 2024.
\newblock URL \url{https://arxiv.org/abs/2407.00079}.

\bibitem[Rajbhandari et~al.(2020)Rajbhandari, Rasley, Ruwase, and He]{rajbhandari2020zero}
Samyam Rajbhandari, Jeff Rasley, Olatunji Ruwase, and Yuxiong He.
\newblock Zero: Memory optimizations toward training trillion parameter models.
\newblock In \emph{SC20: International Conference for High Performance Computing, Networking, Storage and Analysis}, pages 1--16. IEEE, 2020.

\bibitem[Rajbhandari et~al.(2021)Rajbhandari, Ruwase, Rasley, Smith, and He]{rajbhandari2021zero}
Samyam Rajbhandari, Olatunji Ruwase, Jeff Rasley, Shaden Smith, and Yuxiong He.
\newblock Zero-infinity: Breaking the gpu memory wall for extreme scale deep learning.
\newblock In \emph{Proceedings of the International Conference for High Performance Computing, Networking, Storage and Analysis}, SC '21, 2021.

\bibitem[Shoeybi et~al.(2019{\natexlab{a}})Shoeybi, Patwary, Puri, LeGresley, Casper, and Catanzaro]{megatronlm}
Mohammad Shoeybi, Mostofa Patwary, Raul Puri, Patrick LeGresley, Jared Casper, and Bryan Catanzaro.
\newblock Megatron-lm: Training multi-billion parameter language models using model parallelism, 2019{\natexlab{a}}.

\bibitem[Shoeybi et~al.(2019{\natexlab{b}})Shoeybi, Patwary, Puri, LeGresley, Casper, and Catanzaro]{shoeybi2019megatron}
Mohammad Shoeybi, Mostofa Patwary, Raul Puri, Patrick LeGresley, Jared Casper, and Bryan Catanzaro.
\newblock Megatron-lm: Training multi-billion parameter language models using model parallelism.
\newblock \emph{arXiv preprint arXiv:1909.08053}, 2019{\natexlab{b}}.

\bibitem[Team(2025)]{seed2025seed-oss}
ByteDance~Seed Team.
\newblock Seed-oss open-source models.
\newblock \url{https://github.com/ByteDance-Seed/seed-oss}, 2025.

\bibitem[Team and Majumder(2020)]{deepspeed3dblog}
DeepSpeed Team and Rangan Majumder.
\newblock {DeepSpeed}: Extreme-scale model training for everyone.
\newblock \url{https://www.microsoft.com/en-us/research/blog/deepspeed-extreme-scale-model-training-for-everyone/}, 2020.

\bibitem[Team et~al.(2024)Team, Georgiev, Lei, Burnell, Bai, Gulati, Tanzer, Vincent, Pan, Wang, et~al.]{team2024gemini}
Gemini Team, Petko Georgiev, Ving~Ian Lei, Ryan Burnell, Libin Bai, Anmol Gulati, Garrett Tanzer, Damien Vincent, Zhufeng Pan, Shibo Wang, et~al.
\newblock Gemini 1.5: Unlocking multimodal understanding across millions of tokens of context.
\newblock \emph{arXiv preprint arXiv:2403.05530}, 2024.

\bibitem[Vaswani et~al.(2017)Vaswani, Shazeer, Parmar, Uszkoreit, Jones, Gomez, Kaiser, and Polosukhin]{DBLP:journals/corr/attention}
Ashish Vaswani, Noam Shazeer, Niki Parmar, Jakob Uszkoreit, Llion Jones, Aidan~N. Gomez, Lukasz Kaiser, and Illia Polosukhin.
\newblock Attention is all you need.
\newblock \emph{CoRR}, abs/1706.03762, 2017.
\newblock URL \url{http://arxiv.org/abs/1706.03762}.

\bibitem[Wang et~al.(2024)Wang, Qin, Ade~Jacobs, Wu, Holmes, Yao, Rajbhandari, Ruwase, Yang, Yang, and He]{zeroplusplus}
Guanhua Wang, Heyang Qin, Sam Ade~Jacobs, Xiaoxia Wu, Connor Holmes, Zhewei Yao, Samyam Rajbhandari, Olatunji Ruwase, Feng Yang, Lei Yang, and Yuxiong He.
\newblock Zero++: Extremely efficient collective communication for large model training.
\newblock In \emph{ICLR 2024}, March 2024.
\newblock URL \url{https://www.microsoft.com/en-us/research/publication/zero-extremely-efficient-collective-communication-for-large-model-training/}.

\bibitem[Yang et~al.(2024{\natexlab{a}})Yang, Yang, Hui, Zheng, Yu, Zhou, Li, Li, Liu, Huang, Dong, Wei, Lin, Tang, Wang, Yang, Tu, Zhang, Ma, Xu, Zhou, Bai, He, Lin, Dang, Lu, Chen, Yang, Li, Xue, Ni, Zhang, Wang, Peng, Men, Gao, Lin, Wang, Bai, Tan, Zhu, Li, Liu, Ge, Deng, Zhou, Ren, Zhang, Wei, Ren, Fan, Yao, Zhang, Wan, Chu, Liu, Cui, Zhang, and Fan]{qwen2}
An~Yang, Baosong Yang, Binyuan Hui, Bo~Zheng, Bowen Yu, Chang Zhou, Chengpeng Li, Chengyuan Li, Dayiheng Liu, Fei Huang, Guanting Dong, Haoran Wei, Huan Lin, Jialong Tang, Jialin Wang, Jian Yang, Jianhong Tu, Jianwei Zhang, Jianxin Ma, Jin Xu, Jingren Zhou, Jinze Bai, Jinzheng He, Junyang Lin, Kai Dang, Keming Lu, Keqin Chen, Kexin Yang, Mei Li, Mingfeng Xue, Na~Ni, Pei Zhang, Peng Wang, Ru~Peng, Rui Men, Ruize Gao, Runji Lin, Shijie Wang, Shuai Bai, Sinan Tan, Tianhang Zhu, Tianhao Li, Tianyu Liu, Wenbin Ge, Xiaodong Deng, Xiaohuan Zhou, Xingzhang Ren, Xinyu Zhang, Xipin Wei, Xuancheng Ren, Yang Fan, Yang Yao, Yichang Zhang, Yu~Wan, Yunfei Chu, Yuqiong Liu, Zeyu Cui, Zhenru Zhang, and Zhihao Fan.
\newblock Qwen2 technical report.
\newblock \emph{arXiv preprint arXiv:2407.10671}, 2024{\natexlab{a}}.

\bibitem[Yang et~al.(2024{\natexlab{b}})Yang, Yang, Zhang, Hui, Zheng, Yu, Li, Liu, Huang, Wei, Lin, Yang, Tu, Zhang, Yang, Yang, Zhou, Lin, Dang, Lu, Bao, Yang, Yu, Li, Xue, Zhang, Zhu, Men, Lin, Li, Xia, Ren, Ren, Fan, Su, Zhang, Wan, Liu, Cui, Zhang, and Qiu]{qwen2.5}
An~Yang, Baosong Yang, Beichen Zhang, Binyuan Hui, Bo~Zheng, Bowen Yu, Chengyuan Li, Dayiheng Liu, Fei Huang, Haoran Wei, Huan Lin, Jian Yang, Jianhong Tu, Jianwei Zhang, Jianxin Yang, Jiaxi Yang, Jingren Zhou, Junyang Lin, Kai Dang, Keming Lu, Keqin Bao, Kexin Yang, Le~Yu, Mei Li, Mingfeng Xue, Pei Zhang, Qin Zhu, Rui Men, Runji Lin, Tianhao Li, Tingyu Xia, Xingzhang Ren, Xuancheng Ren, Yang Fan, Yang Su, Yichang Zhang, Yu~Wan, Yuqiong Liu, Zeyu Cui, Zhenru Zhang, and Zihan Qiu.
\newblock Qwen2.5 technical report.
\newblock \emph{arXiv preprint arXiv:2412.15115}, 2024{\natexlab{b}}.

\bibitem[Yang et~al.(2025)Yang, Li, Yang, Zhang, Hui, Zheng, Yu, Gao, Huang, Lv, Zheng, Liu, Zhou, Huang, Hu, Ge, Wei, Lin, Tang, Yang, Tu, Zhang, Yang, Yang, Zhou, Zhou, Lin, Dang, Bao, Yang, Yu, Deng, Li, Xue, Li, Zhang, Wang, Zhu, Men, Gao, Liu, Luo, Li, Tang, Yin, Ren, Wang, Zhang, Ren, Fan, Su, Zhang, Zhang, Wan, Liu, Wang, Cui, Zhang, Zhou, and Qiu]{qwen3}
An~Yang, Anfeng Li, Baosong Yang, Beichen Zhang, Binyuan Hui, Bo~Zheng, Bowen Yu, Chang Gao, Chengen Huang, Chenxu Lv, Chujie Zheng, Dayiheng Liu, Fan Zhou, Fei Huang, Feng Hu, Hao Ge, Haoran Wei, Huan Lin, Jialong Tang, Jian Yang, Jianhong Tu, Jianwei Zhang, Jianxin Yang, Jiaxi Yang, Jing Zhou, Jingren Zhou, Junyang Lin, Kai Dang, Keqin Bao, Kexin Yang, Le~Yu, Lianghao Deng, Mei Li, Mingfeng Xue, Mingze Li, Pei Zhang, Peng Wang, Qin Zhu, Rui Men, Ruize Gao, Shixuan Liu, Shuang Luo, Tianhao Li, Tianyi Tang, Wenbiao Yin, Xingzhang Ren, Xinyu Wang, Xinyu Zhang, Xuancheng Ren, Yang Fan, Yang Su, Yichang Zhang, Yinger Zhang, Yu~Wan, Yuqiong Liu, Zekun Wang, Zeyu Cui, Zhenru Zhang, Zhipeng Zhou, and Zihan Qiu.
\newblock Qwen3 technical report.
\newblock \emph{arXiv preprint arXiv:2505.09388}, 2025.

\bibitem[Zhao et~al.(2023)Zhao, Gu, Varma, Luo, Huang, Xu, Wright, Shojanazeri, Ott, Shleifer, Desmaison, Balioglu, Damania, Nguyen, Chauhan, Hao, Mathews, and Li]{fsdp2}
Yanli Zhao, Andrew Gu, Rohan Varma, Liang Luo, Chien-Chin Huang, Min Xu, Less Wright, Hamid Shojanazeri, Myle Ott, Sam Shleifer, Alban Desmaison, Can Balioglu, Pritam Damania, Bernard Nguyen, Geeta Chauhan, Yuchen Hao, Ajit Mathews, and Shen Li.
\newblock Pytorch fsdp: Experiences on scaling fully sharded data parallel.
\newblock \emph{Proc. VLDB Endow.}, 16\penalty0 (12):\penalty0 3848–3860, August 2023.
\newblock ISSN 2150-8097.
\newblock \doi{10.14778/3611540.3611569}.
\newblock URL \url{https://doi.org/10.14778/3611540.3611569}.

\end{thebibliography}

\end{document}